# Discovery of a bright quasar without a massive host galaxy

Pierre Magain[1], Géraldine Letawe[1], Frédéric Courbin[2], Pascale Jablonka[2,3,4], Knud Jahnke[5], Georges Meylan[2] & Lutz Wisotzki[5]

[1]*Institut d'Astrophysique et de Géophysique, Université de Liège, Allée du 6 Août, 17, Bât B5C, B–4000 Liège, Belgium*

[2] *Laboratoire d'Astrophysique, Ecole Polytechnique Fédérale de Lausanne (EPFL), Observatoire, CH–1290 Sauverny, Switzerland*

[3] *Observatoire de l'Université de Genève, CH–1290 Sauverny, Switzerland*

[4] *On leave from GEPI, UMR 8111, Observatoire de Paris, France*

[5]*Astrophysikalisches Institut Potsdam, An der Sternwarte 16, D–14482 Potsdam, Germany*

**Quasars are thought to be powered by the infall of matter onto a supermassive black hole at the centre of massive galaxies[1,2]. As the optical luminosity of quasars exceeds that of their host galaxy, disentangling the two components can be difficult. This led in the 1990's to the controversial claim of the discovery of 'naked' quasars[3–7]. Since then, the connection between quasars and galaxies has been well established[8]. Here we report on the observation of a quasar lying at the edge of a gas cloud, whose size is comparable to that of a small galaxy, but whose spectrum shows no evidence for stars. The gas cloud is excited by the quasar itself. If a host galaxy is present, it is at least six times fainter than would normally be expected[8,9] for such a bright quasar. The quasar is interacting dynamically with a neighbouring galaxy – which matter might be feeding the black hole.**

HE0450–2958 is a bright quasar ($M_V = -25.8$) at a redshift of $z = 0.285$, associated with powerful infrared (IR) emission[10,11]. Early imaging revealed a quasar–galaxy pair with signs of violent dynamical interaction[12], and a starburst in the galaxy. A collision between the two systems about $10^8$ years ago probably triggered both the starburst and the quasar activity[13].

We use the MCS deconvolution technique[14–17] on our new HST images, in order to explore the vicinity of the quasar (Fig 1). These images confirm the strongly irregular shape of the companion galaxy, indicative of gravitational interaction with the quasar. However, the most interesting finding is that no significant host galaxy centered on the quasar position is found – we quantify this statement below. The most prominent feature is a diffuse albeit compact component, just beside the quasar (and to which we shall refer as "the blob" in the following), with a possible very faint extension around the quasar. If interpreted as an image of the host galaxy, we would have the surprising result that the quasar does not reside inside its host, but just next to it. Moreover, with a diameter of no more than 2500 parsec (pc), the blob is comparable in size to M32, the dwarf companion of the Andromeda galaxy M31, but is about 200 times brighter than M32, with an absolute magnitude of $M_V = -22$.

Figure 2 shows a portion of the spatially deconvolved two–dimensional optical spectrum. The full one–dimensional spectra of the quasar, blob, and companion galaxy are displayed on Fig. 3. While the spectrum of the galaxy shows a strong stellar continuum, with low excitation emission lines typical of gas in star–forming regions, the spectrum of the blob does not show any hint of a continuum. It rather consists of a series of emission lines whose intensity ratios indicate excitation by a much harder spectrum than a stellar one. The absence of a continuum component yields the second surprising result: where we would expect stellar light from a host galaxy, we only see an off–centre structure consisting essentially of gas excited by the quasar radiation.

This non–detection of a significant host galaxy around HE0450–2958 may mean that either it is too compact to be separated from the quasar, or that its surface brightness is below our detection limit. These possibilities are investigated by adding to the HST images an artificial elliptical galaxy centered on the quasar position and properly convolved with the HST point spread function (PSF). If we adopt an absolute magnitude of $M_V = -23$, corresponding to the typical host galaxies expected for such a quasar luminosity[8], and if we decrease its size till it cannot be detected anymore in the deconvolved image, we obtain an upper limit of 100 pc for the half–light radius $R_{1/2}$. This is much too small for a quasar host galaxy, for which $R_{1/2}$ typically ranges between 2000 and 15000 pc[8].

In order to set an upper limit on the surface brightness of the putative host, we fix $R_{1/2}$ to 10 kpc, well representative of host galaxies of luminous quasars[8], and we progressively decrease the luminosity until it falls below our detection limit. This gives an upper limit for the stellar population around the quasar, that we discuss below. The absence of any detectable extended structure is confirmed by applying standard PSF subtraction techniques. The sensitivity of our results to the presence of a host galaxy is illustrated by Fig. 4, which shows that, if a typical host galaxy were present around the quasar, we would indeed detect it.

The mass of the central black hole of a quasar of absolute magnitude $M_V = -25.8$ is predicted to be about $8 \times 10^8$ solar masses for an accretion at half the Eddington Limit with a 10% radiative efficiency. This value is in fair agreement with the estimate derived from the width of the $H_\beta$ line[18]. Using the scaling relations recently established between the black hole mass and the host galaxy spheroid luminosity[19], we can compare the predicted magnitude of the host galaxy with our upper limit. For this purpose, we have to make assumptions about the stellar population in the host galaxy. We consider two extreme cases.

On one hand, we assume a smooth spheroid of old stars (10 billion years). The predicted magnitude[20] of the host is then $M_V = -23.0$ and our upper limit $M_V = -21.2$, i.e. five times fainter than expected. On the other hand, in the case of young stars, of an age similar to the starburst population in the companion galaxy (130 million years), the predicted magnitude amounts to $M_V = -23.5$ and our upper limit becomes $M_V = -20.5$, i.e., 16 times fainter. Any reasonable estimate should lie between these two extremes.

As a complementary approach, avoiding uncertainties related to the black hole mass estimates, we compare directly our upper limits to the magnitudes measured for typical quasar host galaxies. We use the HST sample of 17 quasar host galaxies from Floyd et al.[8]. After conversion of their values to our cosmology ($H_0 = 65$ km s$^{-1}$ Mpc$^{-1}$), these quasars have absolute magnitudes in the range $-26.8 < M_V < -23.2$, with an average of $M_V = -24.3$, somewhat fainter than the present quasar ($M_V = -25.8$). A straight line fitting to the host– versus quasar– magnitude relation gives an expected host magnitude of $M_V = -23.2$ for a quasar of $M_V = -25.8$, and the scatter around the mean relation is $\sigma = 0.54$ (Fig. 5). Note that this value of $M_V = -23.2$ nicely falls between the two extreme cases considered in the previous paragraph. Our upper limit is between 6 and 12 times fainter than the expected value, depending on the assumed stellar age. In the most conservative case (i.e. a smooth spheroidal distribution of old stars), the deviation from the average relation is 3.7 $\sigma$, which is highly significant. Moreover, we should point out that such a conservative case is not really expected for galaxies involved in collisions, which generally display a distorted geometry (e.g. tidal tails) and active star formation. Taking these effects into account would make the deviation even more significant.

Finally, the hydrogen emission line ratios, which indicate a strong reddening of the companion galaxy, are close to the theoretically expected values in the quasar and blob. This suggests that obscuration by dust is very limited in the vicinity of the quasar and cannot explain the non–detection of the host galaxy.

This peculiar quasar, therefore, has a host galaxy (if any) that is significantly less luminous than expected from its nuclear luminosity and black hole mass. Moreover, it suffers a strong dynamical interaction with an ultra–luminous infrared galaxy (a rare class of galaxies, systematically involved in collisions[13]). These two peculiarities are most probably related. One might suggest that the host galaxy has disappeared from our view as a result of the collision, but it is hard to imagine how the complete disruption of a galaxy could happen. An alternative suggestion would be that an isolated black hole may have captured gas and become a quasar while crossing the disk of the neighboring galaxy with a low relative speed (the radial velocity difference between the quasar and different parts of the galaxy ranges between –60 and +200 km s$^{-1}$). However, such a gravitational accretion of matter is very inefficient and the dynamical interaction of a large galaxy with an $8\times10^8$ solar masses black hole would probably not induce such strong perturbations as observed in the companion galaxy.

Another possibility would be that the black hole of HE0450–2958 lies in a galaxy with not only a stellar content much lower than average, but also with an important dark halo, a "dark galaxy" [21]. The interaction with such a massive object could more easily explain the peculiarities of the neighboring galaxy as well as the capture of gas, resulting in the ignition of the quasar.

1. Richstone, D., Ajhar, E. A., Bender, R. et al. Supermassive black holes and the evolution of galaxies. *Nature* **395**, A14–A18 (1998)

2. Dunlop, J.S., McLure, R.J., Kukula, M.J., Baum, S. A., O'Dea, C. P. & Hughes, D. H. Quasars, their host galaxies and their central black holes. *Mon. Not. R. astr. Soc.* **340**, 1095–1135 (2003)

3. Bahcall, J.N., Kirhakos, S., Schneider, D.P. HST images of nearby luminous quasars. *Astrophys. J.* **435**, L11–L14 (1994)

This work has been supported by the PPS Science Policy (Belgium), by PRODEX (ESA), and by the Swiss National Science Fundation. The observations were obtained with the ESO/VLT (Paranal, Chile) and with the NASA/ESA Hubble Space Telescope.

Correspondence and requests for materials should be addressed to P. Magain (e–mail: Pierre.Magain@ulg.ac.be).

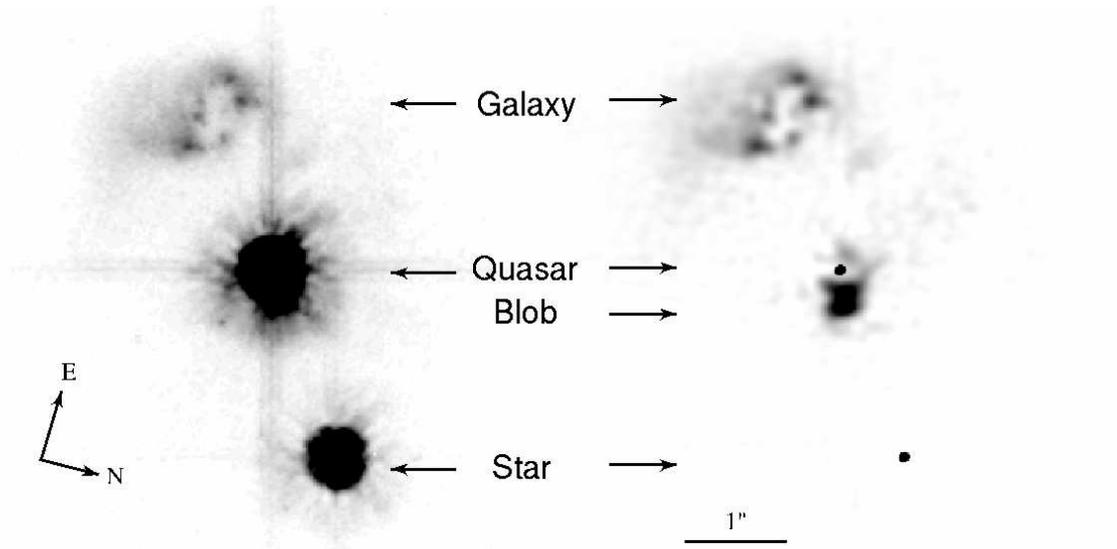

Fig. 1. **HST images of the quasar and its immediate surrounding.** The original HST image of HE0450–2958 is shown on the left side, while the result of the deconvolution with the MCS algorithm is displayed on the right side. The images were obtained on October 1$^{st}$, 2004 with the High Resolution Channel (HRC) of the Advanced Camera for Surveys (ACS) onboard HST. Six dithered exposures of the quasar field were taken through the F606W filter, three short ones (30 s) and three longer ones (330 s). In contrast with previous HST observations of quasar host galaxies, a significant fraction of the observing time was devoted to the PSF characterization, by observing during the same orbit both the nearest bright star and the quasar, always placing the star at the same location as the quasar on the detector. This observational strategy ensures that the PSF temporal and spatial variations are minimized. Because of the extended wings of the PSF, the blob of gas just beside the quasar can only be seen after careful processing of the images. No other related feature is found in the vicinity of the quasar–galaxy pair. From the observed fluxes, and applying the appropriate distance and spectral corrections, we obtain absolute magnitudes of $M_V = -22$ for the blob and $M_V = -23$ for the companion galaxy. After correction for dust extinction, the latter amounts to $M_V = -26$. All magnitudes mentioned in the text have been similarly corrected, and computed with $H_0 = 65$ km s$^{-1}$ Mpc$^{-1}$.

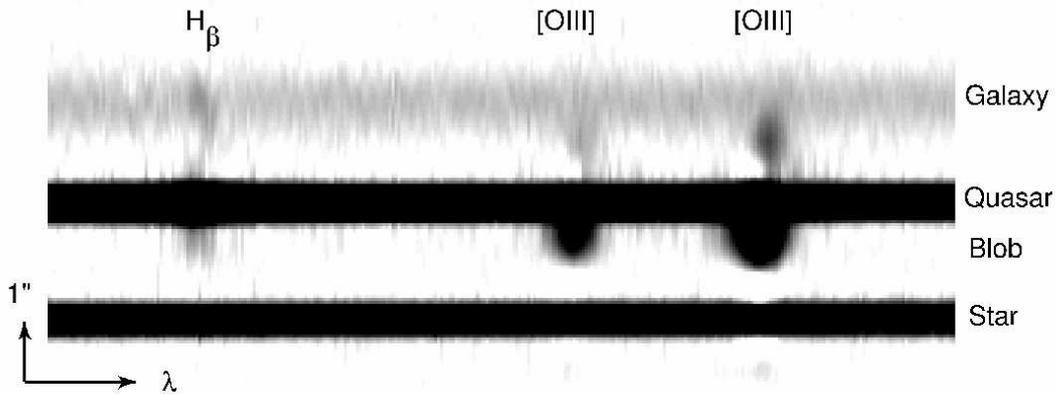

Fig. 2. **Two–dimensional VLT spectra of the quasar and neighbouring objects.** These long slit spectra have been deconvolved in the spatial direction, in order to separate the different objects. The spectral direction is horizontal, while the spatial direction is vertical. The emission lines just below the quasar spectrum are, from left to right, the hydrogen $H_\beta$ (486.1 nm) and the oxygen [O III] (495.9/500.7 nm) doublet corresponding to the blob of ionized gas. Weaker emission lines are also detected in between the quasar and the companion galaxy, showing that some ionized gas is also present there. The vertical arrow gives the spatial scale (1"). Our spectroscopic observations of HE0450–2958 were carried out in a new way, taking advantage of the multi–slit mode of the VLT/FORS1 to record simultaneously the spectra of the quasar and of its host galaxy, as well as the spectra of nearby isolated stars needed for subsequent determination of the PSF. The main slit covers not only the quasar, but also the companion galaxy and a nearby object identified as a foreground star, which is used together with the isolated stars for an accurate determination of the PSF.

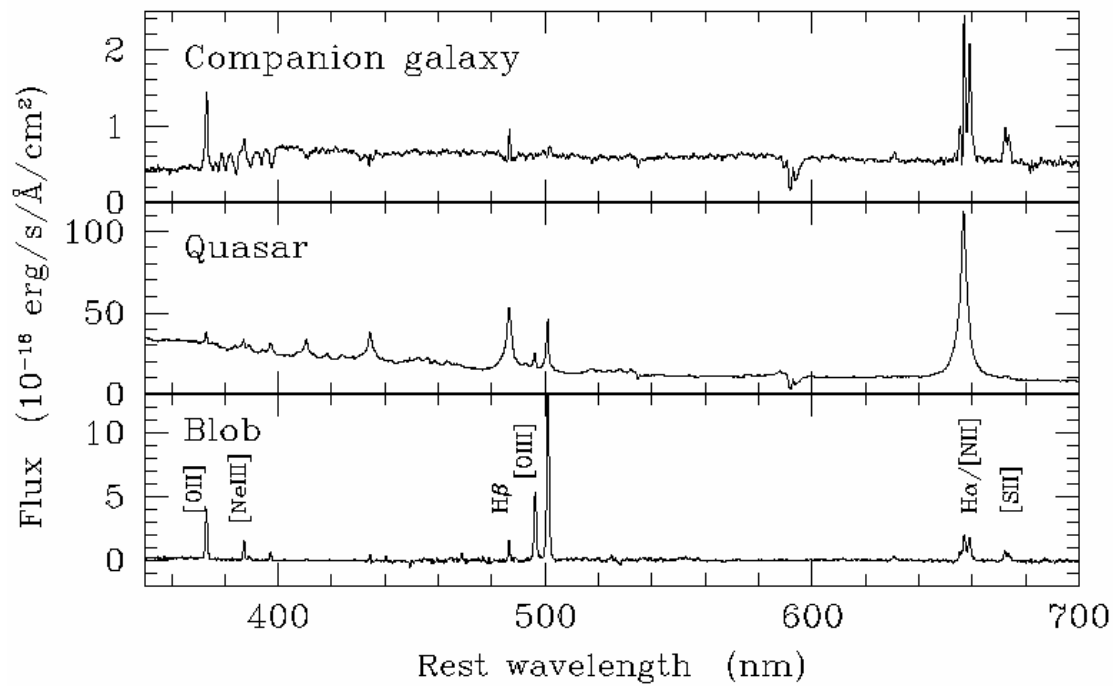

Fig. 3. **Spectra of the companion galaxy, quasar, and blob.** Top: Integrated one–dimensional spectrum of the companion galaxy. Middle: Spectrum of the quasar. Bottom: spectrum of the blob. Note the absence of continuum light in the blob, which would be due to stars, and the emission line ratios (e.g. [O III]/H$_\beta$) indicating ionization by the quasar radiation. The Balmer decrement (H$_\alpha$/H$_\beta$ intensity ratio) indicates that, contrary to the quasar and blob, the companion galaxy is heavily reddened by dust. It is therefore the companion galaxy, and not the quasar, which is the strong IR emitter detected by the IRAS satellite[10]. This also indicates that absorption of light by dust cannot explain the non–detection of a host galaxy around the quasar.

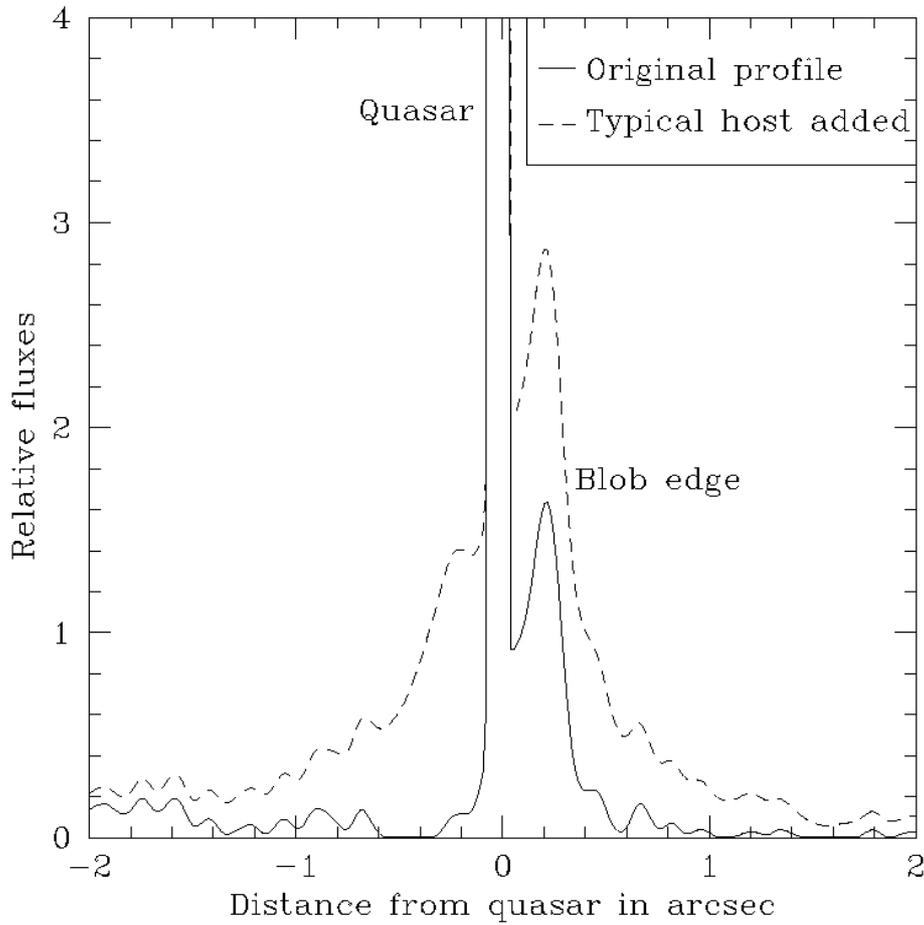

Fig. 4. **Intensity profile through the quasar position.** Solid line: N–S profile of the deconvolved image. The peak just on the right of the quasar corresponds to the edge of the blob. Dashed line: corresponding profile of the deconvolved image after an elliptical host (de Vaucouleurs profile) with $R_{1/2}$ = 10 kpc and $M_V$ = −23 has been artificially added to the original images. The comparison of the two profiles shows that any host galaxy with the expected luminosity would be easily detected.

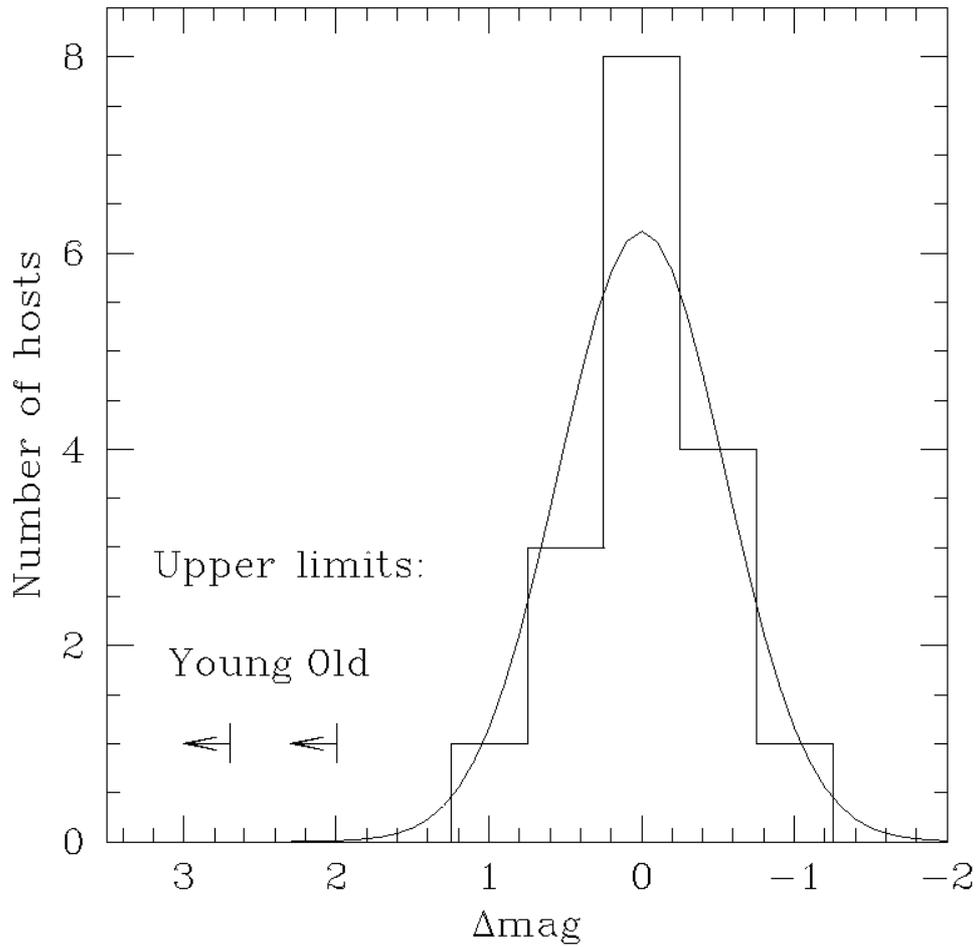

Fig. 5. **Upper limits on the host magnitudes.** The histogram gives the dispersion of the host magnitudes around the mean trend in the HST luminous quasars sample[8]. The curve represents a gaussian fit to these data. Our upper limits for a smooth spheroidal distribution of either old or young stars are indicated. They both deviate from the mean by a large amount of 3.7 and 5 sigma, respectively.